\documentclass[preprint,aps,showpacs,showkeys,preprintnumbers,amsmath,amssymb,floatfix]{revtex4}

\usepackage{graphicx}
\usepackage{dcolumn}
\usepackage{bm}

\def\la{\langle}
\def\ra{\rangle}

\newcommand{\wh}{\widehat}
\newcommand{\nn}{\nonumber}

\newcommand{\FF}{\langle aFF\rangle}

\newcommand{\smvs}{\vbox{\vskip 8mm}}

\newcommand{\qq}{\langle\bar qq\rangle}
\newcommand{\MSb}{{\overline{{\rm MS}}}}

\begin{document}

\preprint{}

\title{ Contribution of $DK$ Continuum in the QCD Sum Rule for
$D_{sJ}(2317)$}

\author{Yuan-Ben Dai}
\email{dyb@itp.ac.cn} \affiliation{Institute of Theoretical Physics,
Chinese Academy of Sciences, P.O. Box 2735, Beijing 100080, China}

\author{Xin-Qiang Li}
\email{xqli@itp.ac.cn} 
\affiliation{Institute of Theoretical Physics, Chinese Academy of
Sciences, P.O. Box 2735, Beijing 100080, China}

\author{Shi-Lin Zhu}
\email{zhusl@th.phy.pku.edu.cn} \affiliation{Department of Physics,
Peking University, Beijing 100871, China}

\author{Ya-Bing Zuo}
\email{yabingzuo@gucas.ac.cn} \affiliation{Department of Physics,
Graduate school of Chinese Academy of Sciences, Beijing, China}

\begin{abstract}
Using the soft-pion theorem and the assumption on the final-state
interactions, we include the contribution of $DK$ continuum into the
QCD sum rules for $D_{sJ}(2317)$ meson. We find that this
contribution can significantly lower the mass and the decay constant
of $D_s(0^+)$ state. For the value of the current quark mass
$m_c(m_c)=1.286~{\rm GeV}$, we obtain the mass of $D_s(0^+)$ $M=2.33
\pm 0.02~{\rm GeV}$ in the interval $s_0=7.5-8.0~{\rm GeV}^2$, being
in agreement with the experimental data, and the vector current
decay constant of $D_s(0^+)$ $f_0=0.128 \pm 0.013~{\rm GeV}$, much
lower than those obtained in previous literature.
\end{abstract}

\pacs{12.39.Hg, 13.25.Hw, 13.25.Ft, 12.38.Lg}

\keywords{Charm-strange mesons, soft-pion theorem.}

\maketitle

\pagenumbering{arabic}

\section{Introduction}
\label{sec1}

In 2003 BaBar Collaboration discovered a positive-parity scalar
charm strange meson $D_{sJ}(2317)$ with a very narrow
width~\cite{babar}, which was confirmed by CLEO~\cite{cleo} later.
In the same experiment CLEO also observed the $1^+$ partner state at
$2460~{\rm MeV}$~\cite{cleo}. Since these two states lie below the
$DK$ and $D^\ast K$ threshold, respectively, the potentially
dominant s-wave decay modes $D_{sJ}(2317) \to DK$ etc., are
kinematically forbidden. Thus the radiative decays and
isospin-violating strong decays become the dominant decay modes.
Therefore both of them are very narrow.

The discovery of these two states has triggered heated discussion on
their nature in literature. The key point is to understand their low
masses. The mass of $D_{sJ}(2317)$ is significantly lower than the
expected values in the range of $2.4-2.6~{\rm GeV}$ within quark
models~\cite{qm}. The model using the heavy-quark mass expansion of
the relativistic Bethe-Salpeter equation predicted a lower value
$2.369~{\rm GeV}$~\cite{jin}, which is still higher than the
experimental data by about $50~{\rm MeV}$.

From the experience with $a_0/f_0(980)$, Van Beveren and
Rupp~\cite{rupp} argued that the low mass of $D_{sJ}(2317)$ could
arise from the mixing between the $0^+$ $\bar{c}s$ state and the
$DK$ continuum. In this way the lowest $0^+$ state could be pushed
much lower than that expected from the quark models.

The mass of $D_s(0^+)$ state from the lattice QCD calculation is
also significantly larger than the experimentally observed mass of
$D_{sJ}(2317)$~\cite{bali,dougall,soni}. It is also pointed out in
Ref.~\cite{bali} that $D_{sJ}(2317)$ might receive a large component
of $DK$ continuum, which makes the lattice simulation very
difficult.

The difficulty with the $\bar{c} s$ interpretation leads many
authors to speculate that $D_{sJ}(2317)$ is a $\bar{c}q s \bar{q}$
four quark state~\cite{cheng,barns}, or a strong $D \pi$
atom~\cite{szc}. However, calculations based on the quark model show
that the mass of the four quark state is much larger than that of
the $0^+$ $\bar{c}s$ state~\cite{vijande,zhang}. The radiative decay
of $D_{sJ}(2317)$ also favors that it is a $\bar{c} s$
state~\cite{colangelo06}. Furthermore, there are two $0^+$ states in
the four quark system and one in the two-quark system. Only one
$0^+$ state has been found below the $2.86~{\rm GeV}$ resonance in
the experimental search by BaBar~\cite{babar06}, consistent with the
$\bar{c} s$ interpretation.

This problem has been treated with QCD sum rules in the heavy quark
effective theory in Ref.~\cite{dai}. The resulting $D_s(0^+)$ mass
is consistent with the experimental data within large theoretical
uncertainties. However, the central value is still larger than the
data by $90~{\rm MeV}$. Even larger result for the $D_s(0^+)$ mass
was obtained in the earlier work with the sum rule in full
QCD~\cite{colangelo91}. It has been pointed out in Ref.~\cite{dai}
that, in the formalism of QCD sum rules, the physics of mixing with
$DK$ continuum resides in the contribution of $DK$ continuum in the
sum rule, and including this characteristic contribution should
render the mass of $D_s(0^+)$ lower.

Recently, there have been two investigations on this problem using
sum rules in full QCD including the $O(\alpha_s)$ corrections. In
Ref.~\cite{haya} the value of the charm quark pole mass
$M_c=1.46~{\rm GeV}$ is used, and the mass of $0^+$ $\bar{c}s$ state
is found to be $100-200~{\rm MeV}$ higher than the experimental
data. On the other hand, in Ref.~\cite{narison} the current quark
mass $m_c=1.15~{\rm GeV}$~(corresponding to $M_c\simeq 1.3~{\rm
GeV}$ to $O(\alpha_s)$) is used, and the central value of the
resulting $0^+$ $\bar{c}s$ mass is in agreement with the data.
However, a low value of $m_c$ is used and the same value of the
continuum threshold (denoted by $s_0$ below) is used for $0^+$
$\bar{c}s$ and  $0^-$ $\bar{c}s$.

On the other hand, the perturbative three loop, order $\alpha_s^2$
correction to the two-point correlation function with one heavy and
one massless quark has been
calculated~\cite{Chetyrkin:2000mq,Chetyrkin:2001je}. It turns out
that in the pole mass scheme used by many previous analyses
including Ref.~\cite{narison, haya}, the perturbative expansion is
far from converging. However, taking the quark mass in the modified
minimal subtraction~($\MSb$) scheme~\cite{Bardeen:1978yd}, better
convergence of the higher order corrections is obtained, and thus a
more reliable determination of physical quantities of the lowest
lying resonances becomes feasible~\cite{Jamin:2001fw}.

Usually the contribution of two-particle continuum is neglected
within the QCD sum rule formalism. However, because of the large
s-wave coupling of $D_s(0^+)DK$~\cite{colangelo95,zhu} and the
adjacency of the $D_s(0^+)$ mass to the $DK$ threshold, this
contribution may not be neglected in the considered case. In the
present work, we shall therefore calculate this contribution and
include it in the QCD sum rule. In the meantime we take into account
the perturbative three loop order $\alpha_s^2$ correction and work
in the $\MSb$ scheme. We find that the $DK$ continuum contribution
indeed renders both the mass and the decay constant of $D_s(0^+)$
significantly lower.

In Section~\ref{sec2}, we give a short overview of the traditional
QCD sum rule for the scalar charm strange meson. Then we derive the
$DK$ continuum contribution and write down the full sum rule in
Section~\ref{sec3}. Finally, the numerical results and our
conclusions are presented in Section~\ref{sec4}. Some relevant
formulas and expressions used in this paper are collected in
Appendices.

\section{The traditional QCD sum rule for the scalar charm-strange meson}
\label{sec2}

We consider the scalar correlation function
\begin{equation}
\Pi(p^2) \; \equiv \; i \int \! dx \, e^{ipx} \, \langle 0 \vert \,
T\{\,j(x)\,j(0)^\dagger\}\vert\ 0 \rangle\,, \label{eq:1.1}
\end{equation}
where the renormalization invariant operator $j(x)$ is defined as
\begin{equation}\label{j}
j(x) \; = \; (m_c-m_s):\!\bar s(x)\,c(x)\!: \,,
\end{equation}
with $m_c$ and $m_s$ being the charm and strange quark current mass,
respectively. Up to a subtraction polynomial in $p^2$, the
correlation function $\Pi(p^2)$ satisfies the following dispersion
relation
\begin{equation}\label{disrel}
\Pi(p^2) \; = \; \int\limits_0^\infty
\frac{\rho(s)}{(s-p^2-i\epsilon)}\,ds + \mbox{subtractions} \,.
\end{equation}

At the quark gluon level, the spectra function $\rho(s)$ is
calculable using the renormalization group improved perturbation
theory in the framework of the operator product expansion~(OPE).
Following Jamin and Lange~\cite{Jamin:2001fw}, in this paper we
shall adopt the $\MSb$ running quark mass scheme rather than the
pole mass one, and take into account both the $O(\alpha_s^2)$ terms
in the perturbation theory obtained in
Ref.~\cite{Chetyrkin:2000mq,Chetyrkin:2001je} and the corrections
from the light quark mass up to order $m_s^4$. In addition we have
included the contribution from the four quark condensation which
affects the final result for $D_s(0^+)$ mass only by a few ${\rm
Mev}$. For convenience, all the relevant expressions for $\rho(s)$
at the quark-gluon level, denoted by $\rho_{\rm QCD}(s)$, are
summarized in Appendix~A.

On the other hand, $\rho(s)$ can be phenomenologically written in
terms of contributions from intermediate hadronic states. Generally,
the spectral density at the hadronic level, denoted by $\rho_{\rm
H}$, is taken to be the pole term of the lowest lying hadronic state
plus the continuum starting from some threshold, with the latter
identified with the QCD continuum
\begin{equation}
\frac{\rho_{\rm H} (t)}{\pi} = f_{0}^2 M^4 \delta (t-M^2) +
\textbf{QCD continuum} \times \theta(t-s_0)\,,
\end{equation}
where $f_{0}$ is the vector current decay constant of $0^+$ $
\bar{c}s$ particle, analogous to $f_\pi=131~{\rm MeV}$. $M$ is the
mass of this particle, and $s_0$ is the continuum threshold above
which the hadronic spectral density is modeled by that at the quark
gluon level. The recent works~\cite{haya,narison} also use the above
ansatz.

After making the Borel transformation to suppress the contribution
of higher excited states and invoking the quark-hadron duality, one
arrives at the sum rule
\begin{equation}\label{sumrule}
\int dt\, \frac{\rho_{\rm H} (t)}{\pi}\, \exp[-\frac{t}{M_B^2}] =
\int \limits_{M_c^2}^{\infty} dt\, \frac{\rho_{\rm QCD}(t)}{\pi}\,
\exp[-\frac{t}{M_B^2}]\,.
\end{equation}
Following Ref.~\cite{Jamin:2001fw}, the lower limit of the
integration in the above equation is taken to be the charm quark
pole mass $M_c$, which can be expressed in terms of the running mass
$m_c(\mu_m)$ through the perturbative three-loop relation as defined
in Appendix~B.

\section{The contribution of $DK$ continuum}
\label{sec3}

The contribution of two-particle continuum to the spectral density
can safely be neglected in many cases, as usually done in the
traditional QCD sum rule analysis. One typical example is the $\rho$
meson sum rule, where the two pion continuum is of p-wave nature.
Its contribution to the spectral density is tiny and the $\rho$ pole
contribution dominates.

However, there may be an exception when the $0^+$ particle couples
strongly to the two-particle continuum via s-wave. In such case,
there is no threshold suppression and the two-particle continuum
contribution may be more significant. The strong coupling of the
$0^+$ particle with the two-particle state and the adjacency of the
$0^+$ mass to the $DK$ continuum threshold result in large coupling
channel effect, which corresponds to the configuration of mixing in
the formalism of quark model. In the problem under consideration,
the mass of $D_{sJ}(2317)$ is only about $45~{\rm MeV}$ below the
$DK$ threshold, and the s-wave coupling of $D_s(0^+)DK$ is found to
be very large~\cite{colangelo95,zhu}. Therefore, one may have to
take into account the $DK$ continuum contribution carefully.

The importance of $D\pi$ continuum contribution in the sum rule for
$D(0^+)$ meson was first emphasized in Ref.~\cite{shifman}, based on
the duality consideration in the case where the $D(0^+)$ mass is
higher than the $D\pi$ threshold. Based on the soft pion theorem,
two of us also made a crude analysis of the $B\pi$ continuum
contribution in the case where the $0^+$ particle mass is higher
than the threshold~\cite{zhu}. In this work, we calculate the
continuum contribution more carefully in the case where the $0^+$
particle mass is lower and very close to the two-particle continuum
threshold.

\begin{figure}[tbh]
\begin{center}
\scalebox{0.9} {\includegraphics{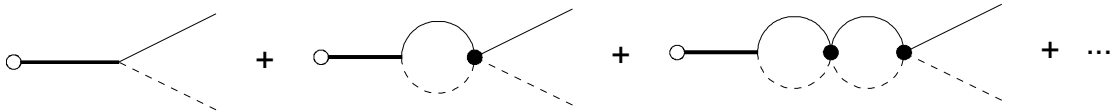}} \caption{Heavy,
light, and dotted lines represent $D_s(0^+)$, $D$, and $K$,
respectively. Black circle represents the Born s-wave amplitude of
$DK$ scattering, and blank one the scalar current.} \label{fig1}
\end{center}
\end{figure}

Let $F(t)$ be the form-factor defined by
\begin{eqnarray}
F(t)=\langle 0| \bar c (0) s(0) |DK\rangle\,.
\end{eqnarray}
From the large s-wave coupling of $D_s(0^+)DK$ and the adjacency of
the $D_s(0^+)$ mass to the $DK$ threshold, one expects that in the
low energy region, $F(t)$ is dominated by the product of a factor of
the $D_s(0^+)$ pole and a factor from the final state interactions.
In the low energy region with $(m_D+m_K)^2 < t < s_0 \leq 8~{\rm
GeV}^2$ needed in our sum rule, the effect of inelastic $DK$
scattering is suppressed by the phase space. Therefore, we take the
approximation to consider only the $DK$ scattering with only elastic
intermediate states. It can be described by the $D_s(0^+)DK$
interaction and the $DDKK$ chiral interaction in the low energy
effective lagrangian, which can be represented by a series of bubble
diagrams shown in Fig.~\ref{fig1}.

The s-wave Born amplitude of $DK$ scattering represented by the
black circles in Fig.~\ref{fig1} contains three terms. The first one
is the $t$ channel pole term $\frac{-ig_0^2}{t-M_0^2}$, with $g_0$
being the $D_s(0^+)DK$ coupling constant and $M_0$ being the mass
parameter normalized at the scale $m_D^2$ in the effective
lagrangian.

The second term corresponds to the direct $DDKK$ interaction in the
effective lagrangian. Let $p,k$ and $p',k'$ be the four momentum of
$D,K$ mesons in the initial and final state respectively, and
$s=(p-k')^2=(p'-k)^2$. In the chiral effective lagrangian, the
amplitudes for the processes $D^+ K^0\rightarrow D^+K^0$, $D^0K^+
\rightarrow D^0K^+$, and $D^+K^0 \leftrightarrow D^0K^+$ in the low
energy $k_0$ region of $K$ meson needed in the QCD sum rule are all
equal to
\begin{equation}
\frac{i}{2}\int\limits_{-1}^{+1} {d \cos \theta}\,\frac{t-s}{2f_K^2}
= \frac{i}{2f_K^2} \Big[ 2\sqrt{t}(k_0+k'_0)-2k_0k'_0-k^2-k'^2
\Big]\,,
\end{equation}
in the center of mass system. Here $k$ and $k'$ should be separately
included in the integrals for two adjacent loops.

The third term of the s-wave Born amplitude arises from the crossing
$s$ channel pole term
\begin{eqnarray}
-i\frac{g_0^2}{2} \int\limits_{-1}^{+1} {d\cos \theta}\,
\frac{1}{s-M_0^2}
=i\frac{g_0^2}{2}\,\frac{1}{B}\,\ln{\frac{A-B}{A+B}}\,,
\end{eqnarray}
where
\begin{equation}
A=t-2\sqrt{t}(k_0+k'_0)+2k_0k'_0+2m_K^2-M_0^2, \qquad
B=2|\vec{k}||\vec{k'}|\,.
\end{equation}
For simplicity, we put the on-shell values of $k_0$, $k'_0$,
$|\vec{k}|$, $|\vec{k'}|$ into $A$ and $B$ in the above equations.
The effect of this approximation on our final results is expected to
be small, since the contribution of the $s$ channel pole term is
relatively small. As a result, one finds that the $s$ channel pole
term is an analytic function of $t$ with only a short cut, the
length of which is only $0.146~{\rm GeV}^2$ for experimental values
of the corresponding masses. Therefore, it can be well approximated
by a pole form $-i\frac{c g_0^2}{t-t_0}$, where
\begin{eqnarray}
&& t_0={{1\over 2}\,
\left[\,2m_D^2+2m_K^2-M_{0}^2+\frac{(m_D^2-m_K^2)^2}{M_{0}^2}\,\right]}
\,,
\end{eqnarray}
\begin{eqnarray}
c={{2m_D^2+2m_K^2-M_{0}^2-{(m_D^2-m_K^2)^2\over
M_{0}^2}}\over{{(m_D^2-m_K^2)^2\over t_0} +t_0-2m_D^2-2m_K^2}} \,.
\end{eqnarray}

With the above results for the three terms of the s-wave Born
amplitude, we can now evaluate the sum of the series of the bubble
diagrams shown in Fig.~\ref{fig1}. Let $f_n(t)$ be the partial sum
of the series of loop diagrams in Fig.~\ref{fig1}, with the loop
number less or equal to $n$. It can then be written in the form
\begin{equation}\label{fnt}
f_n(t)=\frac{-1}{2f_K^2}\,\bigg\{\Big[(2\sqrt{t}\,k'_0-k'^2) -
2f_K^2(\frac{g_0^2}{t-M_0^2}+\frac{c g_0^2}{t-t_0})\Big]\,f_{n0} +
2\,(\sqrt{t}-k'_0)\,f_{n1} -f_{n2}\bigg\}\,,
\end{equation}
where $k'_0$ and $k'$ is the energy and momentum of final-state $K$
meson, and the three unknown functions $f_{ni}$~($i=0,1,2$)
correspond to the diagrams with a factor $1$, $k_0$, and $k^2$
respectively at the last vertex, which contributes to the
integration over the last loop of each diagram.

Let $\Sigma_i(t)$ be integrals defined by Eqs.(C1)-(C6) which appear
as the loop integrals of the individual loop diagrams shown in
Fig.~\ref{fig1}. They can be evaluated using dimensional
regularization~\cite{DR}, with the corresponding analytic results
given in (C1)-(C6). A recurrence relations can be written between
$f_{n}(t)$ and $f_{(n+1)}(t)$, and hence between $f_{ni}(t)$ and
$f_{(n+1)i}(t)$, the coefficients of which are linear combinations
of the loop-integral functions $\Sigma_i(t)$. Taking the limit
$\lim_{n\rightarrow\infty}f_{ni}(t)=f_i(t)$,
$\lim_{n\rightarrow\infty}f_{n}(t)=f(t)$ and separating out terms
with the factor $1$, $k'_0$, and $k'^2$, we can obtain a system of
three linear equations for the three unknown functions $f_i(t)$
\begin{eqnarray}\label{eq1}
&& 2\,f_1\,\Sigma_4 + f_0\,\Sigma_5 - 2\,\Big[ f_1\,\left(
2\,\Sigma_2 + \Sigma_3 \right)  + 2\,f_0\,\Sigma_4 \Big]
\,{\sqrt{t}} + 4\,\left( f_1\,\Sigma_1 + f_0\,\Sigma_2 \right)\,
t \nonumber \\
&& + 4\,f_K^4\,g_0^2\,\Big[ g_0\,\Sigma_0 - f_0\,\left( M_0^2 +
g_0^2\,\Sigma_0 - t \right) \Big]\, \Big[ \frac{c}{{\left( t - t_0
\right) }^2} + \frac{1}{\left( t - M_0^2 \right) \,\left( t - t_0
\right) } \Big]\, \nonumber \\
&& + 2\,f_K^2\,\frac{g_0\,\left( 1 - 2\,f_0\,g_0 \right) \,\left(
2\,\Sigma_1\,{\sqrt{t}} - \Sigma_3 \right)  + 2\,f_1\,\Big[
g_0^2\,\Sigma_1 - \left( M_0^2 + g_0^2\,\Sigma_0
\right) \,{\sqrt{t}} + t^{\frac{3}{2}} \Big]}{t - M_0^2}\,\nonumber \\
&& + f_2\,\bigg[\Sigma_3 - 2\,\Sigma_1\,{\sqrt{t}} + 2\, f_K^2\,
g_0^2\,\Sigma_0\,\left( \frac{1}{t-M_0^2} + \frac{c}{t - t_0}
\right) - 2\, f_K^2 \bigg]\, \nonumber \\
&& + 2\,f_K^2\, \frac{2\,c\,g_0^2\, \Big[ f_1\,\Sigma_1 +
f_0\,\Sigma_3 - \left( f_1\,\Sigma_0 + 2\,f_0\,\Sigma_1 \right)
\,{\sqrt{t}} \Big]\,}{t - t_0}\, \nonumber \\
&& + 4\,c\,f_0\,f_K^4\,g_0^4\,\Sigma_0\, \Big[ \frac{c}{{\left( t -
t_0 \right) }^2} + \frac{1}{\left( t - M_0^2 \right) \,\left( t -
t_0 \right) } \Big]=0 \,,
\end{eqnarray}
\begin{eqnarray}\label{eq2}
&& f_2\,\Sigma_1 + f_0\,\Sigma_4 - \Big[ f_2\,\Sigma_0 + f_0\,\left(
2\,\Sigma_2 + \Sigma_3 \right) \Big] \,{\sqrt{t}} + 2\,f_1\,\left(
-f_K^2 + \Sigma_2 - 2\,\Sigma_1\,{\sqrt{t}} + \Sigma_0\,t \right)\,
\nonumber \\
&& + 2\,f_K^2\,\frac{g_0\,\left( 1 - f_0\,g_0 \right) \,\Sigma_1 +
\Big[ f_0\,M_0^2 - g_0\,\left( 1 - f_0\,g_0 \right) \,\Sigma_0 \Big]
\,{\sqrt{t}} - f_0\,t^{\frac{3}{2}}}{M_0^2 - t} +
2\,f_0\,\Sigma_1\,t \nonumber \\
&& + 2\,f_K^2\, \frac{c\,f_0\,g_0^2\,\left(\Sigma_1 -
\Sigma_0\,{\sqrt{t}} \right) } {t - t_0}=0\,,
\end{eqnarray}
\begin{eqnarray}\label{eq3}
&& 2\,f_1\,\Sigma_1 + f_0\,\bigg[\Sigma_3 - 2\,\Sigma_1\,{\sqrt{t}}
+ 2\, f_K^2\, g_0^2\,\Sigma_0\,\left( \frac{1}{t-M_0^2} + \frac{c}{t
- t_0} \right) - 2\, f_K^2 \bigg]\, \nonumber \\
&& + \Sigma_0\,\left( f_2 - \frac{2\,f_K^2\,g_0}{t-M_0^2} -
2\,f_1\,{\sqrt{t}} \right)=0 \,,
\end{eqnarray}
from which the analytic forms for $f_i(t)$ can then be deduced. The
results are shown in Eqs. (C9)-(C12).

Finally, with the explicit expressions for $f_i(t)$ and
$\Sigma_i(t)$ given in Appendix~C, and putting the on-shell value of
$k'$ to Eq.~(\ref{fnt}), we obtain
\begin{eqnarray}\label{ft}
F(t) &=& \frac{-g_0}{t-M_0^2} -
\frac{1}{2f_K^2}\,\bigg\{\Big[(t-m_D^2) -
2f_K^2(\frac{g_0^2}{t-M_0^2}+\frac{c g_0^2}{t-t_0})\Big]\,f_{0} +
\frac{m_D^2-m_K^2}{\sqrt{t}}\,f_{1}
-f_{2}\bigg\} \nonumber \\
&=& {{\lambda} \over {t-M_0^2-\Delta(t)}}\,,
\end{eqnarray}
where $\Delta(t)$ is given by Eq.~(\ref{deltat}).

Similarly, with the same series of $DK$ loops included, the full
propagator of $D_s(0^+)$ meson is related to the function $f_0(t)$
through
\begin{equation}
{\rm Prop(t)}=\frac{i}{t-M_0^2}\,\big[1-g_0f_0(t)\big]\,.
\end{equation}
Using the solution for $f_0(t)$ obtained above and given by
Eq.~(\ref{fit}), it can be further rewritten as
\begin{equation}
{\rm Prop(t)}=\frac{1}{t-M_0^2-\Delta_1(t)}\,,
\end{equation}
with $\Delta_1(t)$ given by Eq.~(\ref{delta1t}).

We have chosen the scale $\mu$ so that the mass parameter $M_0$ is
the physical mass of $D_s(0^+)$ in our approximation, i.e.,
$\Delta_1(M_0^2)=0$. The bare coupling constant $g_0$ is related to
the physical coupling constant $g$ by $g=g_0/\sqrt{Z}$, where
\begin{eqnarray}
Z={d\over {dt}}\left[\,t-M_0^2-\Sigma(t)\,\right]_{t=M_0^2}\,,
\end{eqnarray}
is the on-shell wave function renormalization constant of $D_s(0^+)$
meson.

In order to fix the unknown constant $\lambda$ in $F(t)$ given by
Eq.~(\ref{ft}), we apply the soft-pion theorem
\begin{eqnarray}
F(m_D^2)=\frac{f_D m_D^2}{f_K (m_c+m_u)}
\end{eqnarray}
to the extrapolated  value of the matrix element $\langle 0| \bar c
(0) s(0) |DK\rangle$ at $t=m_D^2$,  from which we can deduce the
constant
\begin{eqnarray}
\lambda =\frac{f_D m_D^2}{f_K (m_c+m_u)}\,
\big[M^2_D-M_0^2-\Delta(m_D^2)\big]\,.
\end{eqnarray}

With all the above equipments, the $DK$ continuum contribution to
the hadronic spectral function can then be written as
\begin{eqnarray}\label{rhoDK}
\rho_{DK}(t) &\;=\;& \frac{1}{8\pi^2} \,\sqrt{1-{(m_D+m_K)^2\over
t}}\, \sqrt{1-{(m_D-m_K)^2\over t}}\,(m_c-m_s)^2\, |F(t)|^2 \nn \\
&& \hspace{-0mm} \times \theta(\sqrt{t}-m_D-m_K)\,\theta(s_0-t) \,.
\end{eqnarray}

In the above calculations we have neglected the contribution of
the $D_s\eta$ channel. The threshold of this channel is at
$t=6.329~{\rm GeV}^2$. Our formula for the contribution of the
two-particle term is proportional to $(t-M_0^2)^{-2}$. The lower
part of the integration in $t$ is more important. At the
thresholds of the two channels the factor $(t-M_0^2)^{-2}$ for the
$DK$ channel is about 18 times larger than that for the $D_s\eta$
channel. Therefore, the effect of the latter is expected to be
small.

\section{Numerical results and discussions}
\label{sec4}

In our numerical analysis, we use the recent result for c quark
current mass $m_c(m_c)=1.286~{\rm GeV}$~\cite{Kuhn:2007vp}. Other
input parameters are the following ~(assuming the isospin symmetry):
$\alpha_s(m_Z)=0.1189$~\cite{Bethke:2006ac}, $m_s(2~{\rm
GeV})=96.10~{\rm MeV}$~\cite{Narison:2005ny}, $m_u(2~{\rm
GeV})=m_s(2~{\rm GeV})/24.4$~\cite{Narison:2005ny}, $\langle \bar s
s\rangle =0.8\times(-0.243)^3~{\rm GeV}^3$~\cite{narison}, $\langle
\bar s g_s \sigma \cdot G s\rangle =0.8~{\rm GeV}^2\times\langle
\bar s s\rangle$~\cite{narison}, $\langle \alpha_s G^2\rangle
=0.06~{\rm GeV}^4$~\cite{narison}, in the four quark condensation
term Eq.~(\ref{psibarpsi0}) $\sigma=3$,
$m_D=\frac{m_{D^{\pm}}+m_{D^0}}{2}=1866.9~{\rm MeV}$~\cite{pdg},
$m_K=\frac{m_{K^{\pm}}+m_{K^0}}{2}=495.66~{\rm MeV}$~\cite{pdg},
$f_D=222.6~{\rm MeV}$~\cite{pdg}, $f_K=159.8~{\rm MeV}$~\cite{pdg}.

\begin{figure}[tbh]
\begin{center}
\scalebox{0.8} {\includegraphics{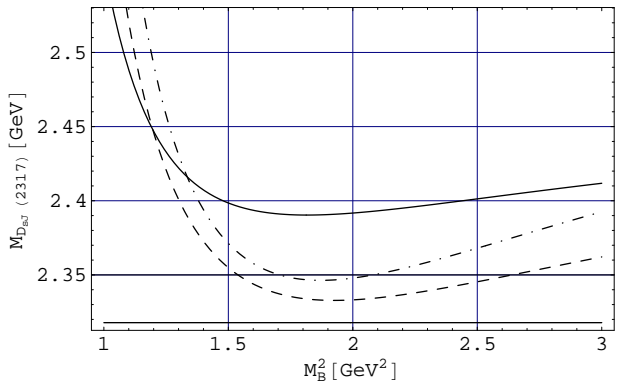}} \caption{The variation
of $M$ with $M_B^2$ when $s_0=8.0~{\rm GeV}^2$. The solid,
dashdotted, and dashed curves are for the case without the $DK$
continuum contribution, $g=4.0~{\rm GeV}$, and $g=7.0~{\rm GeV}$,
respectively.} \label{fig2}
\end{center}
\end{figure}

The renormalized coupling constant $g$ was determined to be in the
interval $g=5.1-7.5~{\rm GeV}$ in Refs.~\cite{colangelo95,zhu}.
Inclusion of the contribution of $DK$ continuum in the sum rule
analysis of the scalar current channel will lower the $g$ value.
Since the uncertainty is large, we have not calculated this
correction and simply allow the renormalized $g$ to vary in the
region $g=4.0-7.0~{\rm GeV}$.

\begin{figure}[tbh]
\begin{center}
\scalebox{0.8} {\includegraphics{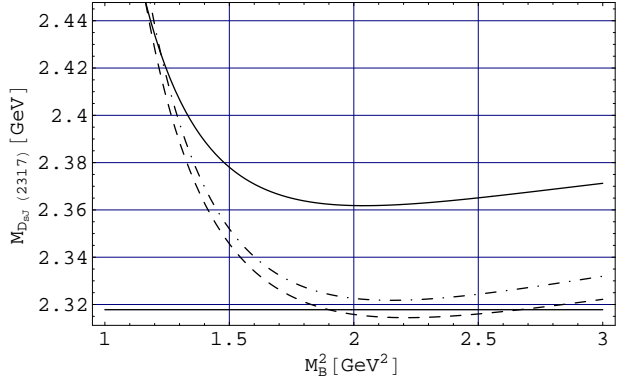}} \caption{The variation
of $M$ with $M_B^2$ when $s_0=7.5~{\rm GeV}^2$. The other captions
are the same as in Fig.~\ref{fig2}.} \label{fig3}
\end{center}
\end{figure}

A resonance of the $D_s$ system with the natural parity has has
been observed experimentally at $t=8.18~{\rm
GeV}^2$~\cite{babar06}. If it is an excited state of $D_s(0^+)$,
we should confine us to $s_0$ smaller than and close to $8.0~{\rm
GeV}^2$. We shall first consider this case and then discuss the
case that this resonance is not a $0^+$ state. The convergence of
the OPE series and dominance of the sum by the pole and the $DK$
continuum terms over the QCD continuum beyond $s_0$ constrain the
Borel mass $M_B$ in a region depending on the parameters
$m_c(m_c)$ and $s_0$. Taking $m_c=1.286~{\rm GeV}$, $M_B^2 \in
[0.99, 2.68]~{\rm GeV}^2$ for $s_0=8.0~{\rm GeV}^2$, and $M_B^2\in
[0.99, 2.41]~{\rm GeV}^2$ for $s_0=7.5~{\rm GeV}^2$. As mentioned
in Refs.~\cite{Jamin:2001fw,Bordes:2005wi,Bordes:2004vu}, the
convergence of the perturbative expansion of the two-point
correlation function, when written in terms of the pole quark
mass, is rather poor, the order $\alpha_s$ and $\alpha_s^2$ loop
contributions being of similar size with, or even larger than, the
leading term, while the expansion in terms of the $\MSb$ running
mass converges much faster. However, it should be noted that, even
in the $\MSb$ running mass scheme, the convergence of the
asymptotic series in the $D_s$ meson system is worse than the one
found in the $B_s$ meson system. For $D_s(0^+)$ the first order
correction amounts to about $53\%$ and the second order to about
$47\%$ of the leading term using the values of our input
parameters and $s_0=8.0~{\rm GeV}^2$. The same observation has
also be made in Ref.~\cite{Bordes:2005wi}.

We first move the $DK$ continuum contribution to the right hand side
of the sum rule. Then we obtain the curve of $M$ with respect to
$M_B$ by taking the derivative of the logarithm of both sides of the
sum rule as usually done. Since this curve depends on the unknown
parameters $M_0$, we have to do it self-consistently by requiring
that the $M$ value determined by the sum rule for the input ``trial"
value of $M_0$ both lies in the middle of the stability window and
equals roughly to $M_0$. For reliability of the results we also
require that the ratio of $DK$ contribution to the whole sum rule is
not larger than about $60\%$.

\begin{figure}[tbh]
\begin{center}
\scalebox{0.8} {\includegraphics{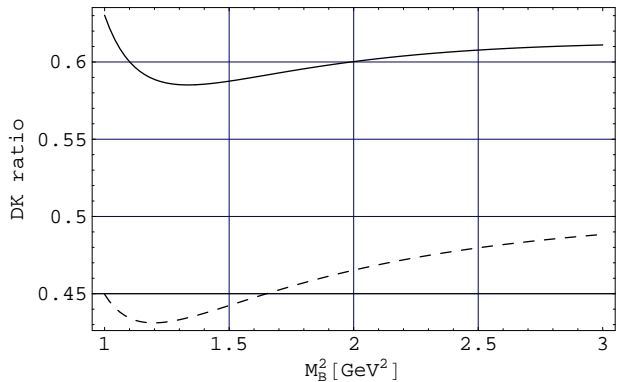}} \caption{The ratio of
the $DK$ continuum contribution as a function of $M_B^2$ with $s_0
=8.0~{\rm GeV}^2$. The solid and dashed curves are for $g=4.0~{\rm
GeV}$, and $g=7.0~{\rm GeV}$, respectively.} \label{fig4}
\end{center}
\end{figure}

\begin{figure}[tbh]
\begin{center}
\scalebox{0.8}{\includegraphics{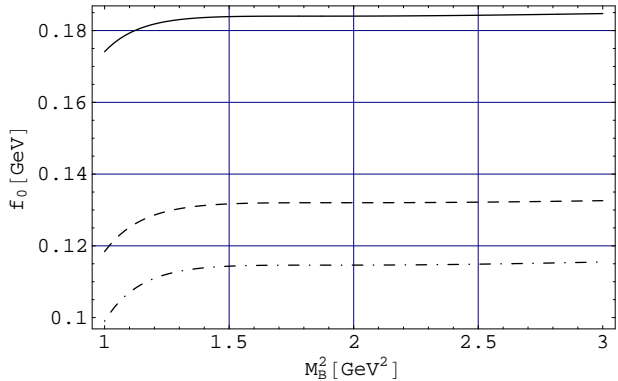}} \caption{The vector current
decay constant $f_0$ as a function of $M_B^2$ with $s_0=8.0~{\rm
GeV}^2$. The other captions are the same as in Fig.~\ref{fig2}.}
\label{fig5}
\end{center}
\end{figure}

With the input $m_c(m_c)=1.286~{\rm GeV}$, we present the
variation of $M$ with $M_B^2$ for $s_0=8.0~{\rm GeV}^2$ and
$s_0=7.5~{\rm GeV}^2$ in Figs.~\ref{fig2} and \ref{fig3},
respectively. For comparison, we also show the case without $DK$
continuum contribution with the same set of input parameters. It
can be seen clearly from the two figures that the inclusion of the
$DK$ continuum contribution can lower the $M$ value by $60-40~{\rm
MeV}$. The $DK$ continuum contributes around $45\%$ to $60\%$ of
the right hand side of the final sum rule for $s_0=8.0~{\rm
GeV}^2$ as shown in Fig.~\ref{fig4}. Another interesting point is
about the vector current decay constant $f_0$ of $D_s(0^+)$ meson.
We find that the inclusion of $DK$ continuum contribution lowers
the decay constant $f_0$ from about $0.185~{\rm GeV}$ to
$0.115-0.132~{\rm GeV}$ for the same $s_0$ value as can be seen
from Fig.~\ref{fig5}.

For $m_c(m_c)=1.286~{\rm GeV}$ and $s_0=7.5-8.0~{\rm GeV}^2$, we
found $M=2.331 \pm 0.016~{\rm GeV}$, being in agreement with the
experimental data $2317.8 \pm 0.6~{\rm MeV}$~\cite{pdg}. For the
same value of input parameters, we found $f_0=0.128 \pm 0.013~{\rm
GeV}$, which is, however, significantly lower than the ones obtained
in previous literature. Here we have not included the errors due to
uncertainties in the QCD sum rule except those from the variation of
the results in the stability window and the $s_0$ interval, since
our main interest is the central value of the results. The previous
results already shew that the $D_s(0^+)$ mass lies in the large
uncertainty interval of the QCD sum rule~\cite{haya,narison}.

Now we consider the case that the new resonance found in
~\cite{babar06} is not a $0^+$ state. In this case the $s_0$ value
can only be determined by stability analysis. The results for the
mass $M$ found for $s_0=8.5,8.0,7.5,7.0~{\rm GeV}^2$ for $g=7~{\rm
 GeV}$ and $g=4~{\rm GeV}$ are shown in Fig. 6 and Fig. 7 respectively.
The working region for $s_0=8.5~{\rm GeV}^2$ and $s_0=7.0~{\rm
GeV}^2$ are $[0.99,2.96]~{\rm GeV}^2$ and $[0.99,2.13]~{\rm
GeV}^2$ respectively. The region of the $s_0$ value can be chosen
by requiring the least sensitivity of the results for the mass to
the value of $s_0$. It is clear from these figures that this is
the region between $s_0=7.5~{\rm GeV}^2$ and $s_0=8.0~{\rm GeV}^2$
which is just the region chosen above for the case of a $0^+$
resonance at $t=8.18~{\rm GeV}^2$. Also the best stability with
respect to $M_B$ is achieved at $s_0=7.5~{\rm GeV}^2$. Therefore,
the results obtained above are essentially unchanged.

\begin{figure}[tbh]
\begin{center}
\scalebox{0.8} {\includegraphics{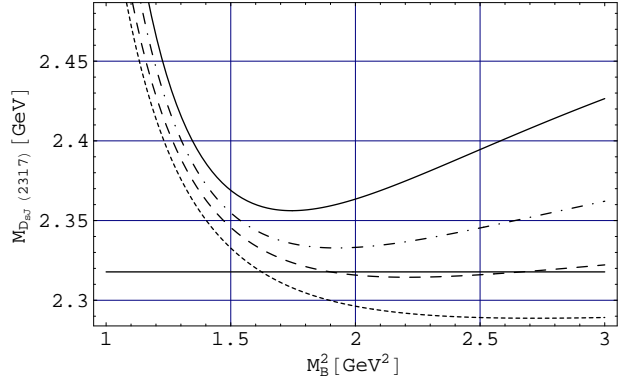}} \caption{The
variation of $M$ with $M_B^2$ when $g=7~{\rm GeV}^2$. The solid,
dashdotted, dashed and dotted curves are for the case $s_0=8.5~{\rm
GeV}^2$, $s_0=8.0~{\rm GeV}^2$,$s_0=7.5~{\rm GeV}^2$,and
$s_0=7.0~{\rm GeV}^2$ respectively.} \label{fig6}
\end{center}
\end{figure}

\begin{figure}[tbh]
\begin{center}
\scalebox{0.8} {\includegraphics{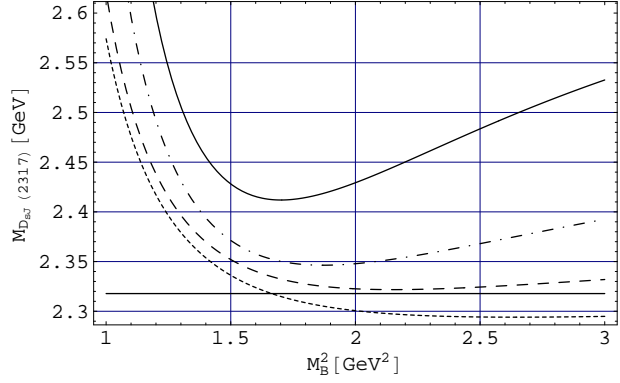}} \caption{The
variation of $M$ with $M_B^2$ when $g=4~{\rm GeV}^2$. The solid,
dashdotted, dashed and dotted curves are for the case $s_0=8.5~{\rm
 GeV}^2$, $s_0=8.0~{\rm GeV}^2$,$s_0=7.5~{\rm GeV}^2$,and
$s_0=7.0~{\rm GeV}^2$ respectively.} \label{fig7}
\end{center}
\end{figure}

The above results show that the contribution of $DK$ continuum,
which contains the physics of the coupled channel effect in the
formalism of QCD sum rule, is significant and is partly the reason
for the unexpected low mass of $0^+$ $\bar c s$ state. Our analysis
also explains partly why the extracted mass of the $0^+$ $\bar c s$
state from the quenched lattice QCD simulation is higher than the
experimental value where the DK continuum contribution was not
included.

\vspace{0.5cm} \noindent {\bf Acknowledgements} We thank Prof.
H.-Y. Cheng for beneficial discussion and Dr Chun Liu and Dr.
Xiang Liu for their kind help. S.L.Z. was supported by the
National Natural Science Foundation of China under Grants 10625521
and 10721063 and Ministry of Education of China.

\newpage
\appendix{\LARGE\bf Appendices}

\section{The spectral function $\rho_{\rm QCD}(s)$ at the quark gluon level}

In this appendix, all the relevant expressions for the spectral
function $\rho(s)$ are given. For further details, we refer the
readers to Ref.~\cite{Jamin:2001fw} and references therein.

\subsection{The perturbative spectral function}

In perturbation theory, the spectral function $\rho_{\rm QCD}(s)$
has an expansion in powers of the strong coupling constant
\begin{equation}\label{rhopt}
\rho_{\rm QCD}(s) = \rho^{(0)}(s) + \rho^{(1)}(s)\,a(\mu_a) +
\rho^{(2)}(s)\,a(\mu_a)^2 + \ldots \,,
\end{equation}
with $a(\mu_a) \equiv \alpha_s(\mu_a)/\pi$. The leading order term
$\rho^{(0)}(s)$ results from a calculation of the bare
quark-antiquark loop and is given by
\begin{equation}\label{rho0}
\rho^{(0)}(s) =
\frac{N_c}{8\pi^2}\,(m_c+m_s)^2\,s\,(1-\frac{m_c^2}{s})^2 \,,
\end{equation}
and, up to order $m_s^4$, the corrections in small mass $m_s$ can be
found in Ref.~\cite{Jamin:1992se}
\begin{equation}\label{rho0m}
\rho^{(0)}_m(s) = \frac{N_c}{8\pi^2}\,(m_c+m_s)^2\,\left\{ 2(1-x)m_c
m_s - 2 m_s^2 - 2\,\frac{(1+x)}{(1-x)}\frac{m_c m_s^3}{s} +
\frac{(1-2x-x^2)}{(1-x)^2} \frac{m_s^4}{s} \,\right\} \,,
\end{equation}
where $x \equiv m_c^2/s$, and the appearing quark masses correspond
to the running masses in the $\MSb$ scheme with $m_c(\mu_m)$ and
$m_s(\mu_m)$ evaluated at the scale $\mu_m$

The order $\alpha_s$ correction $\rho^{(1)}(s)$ can be written as
\begin{eqnarray}\label{rho1}
\rho^{(1)}(s) &\!=\!& \frac{N_c}{16\pi^2}\,C_F\,
(m_c+m_s)^2\,s\,(1-x) \biggl\{(1-x)\Big[\,4L_2(x)+2\ln
x\ln(1-x)-\,(5-2x) \nn \\ \smvs &&\hspace{-0mm} \times
\ln(1-x)\,\Big] + (1-2x)(3-x)\ln x + 3(1-3x)\ln
\frac{\mu_m^2}{m_c^2} + \frac{1}{2}(17-33x)\,\biggr\} \,,
\end{eqnarray}
where $L_2(x)$ is the dilogarithmic function. The order $\alpha_s$
mass corrections to the spectral function can be obtained by
expanding the results given by
\cite{Broadhurst:1981jk,Generalis:1990id} up to order $m_s^4$, after
the higher dimensional operators have been expressed in terms of
non-normal ordered condensates
\begin{eqnarray}\label{rho1m1}
\rho^{(1)}_m(s) &=& \frac{N_c}{8\pi^2}\,C_F\, (m_c+m_s)^2\, m_c
m_s\,\biggl\{(1-x)\Big[\,4L_2(x)+2\ln x\ln(1-x) -\,2(4-x) \nn \\
\smvs &&\hspace{-0mm} \times \ln(1-x)\,\Big] + 2(3-5x+x^2)\ln x
+ 3(2-3x)\ln \frac{\mu_m^2}{m_c^2} + 2(7-9x)\,\biggr\} \,, \\
\vbox{\vskip 12mm} \label{rho1m2} \rho^{(1)}_{m^2}(s) &=&
-\,\frac{N_c}{8\pi^2}\,C_F\, (m_c+m_s)^2\,m_s^2\,
\biggl\{(1-x)\Big[\,4L_2(x)+2\ln x\ln(1-x)\,\Big] \nn \\
\smvs &&\hspace{-0mm} -\,(2+x)(4-x)\ln(1-x) + (6+2x-x^2)\ln x + 6\ln
\frac{\mu_m^2}{m_c^2} + (8-3x)\,\biggr\} \,, \\
\vbox{\vskip 12mm} \label{rho1m3}
\rho^{(1)}_{m^3}(s) &=&
-\,\frac{N_c}{8\pi^2}\,C_F\, (m_c+m_s)^2\,\frac{m_c m_s^3}
{s}\,\biggl\{\,4L_2(x)+2\ln x\ln(1-x) + \frac{(9+8x-9x^2)}{(1-x)^2} \nn \\
\smvs &&\hspace{-0mm} -\,2\frac{(7+7x-2x^2)}{(1-x)}\ln(1-x) +
2\frac{(6+7x-2x^2)} {(1-x)}\ln x +
6\frac{(2-x^2)}{(1-x)^2}\ln\frac{\mu_m^2} {m_c^2}\,\biggr\} \,, \\
\vbox{\vskip 12mm} \label{rho1m4}
\rho^{(1)}_{m^4}(s) &=&
\frac{N_c}{8\pi^2}\,C_F\, (m_c+m_s)^2\,
\frac{m_s^4}{s}\,\biggl\{\,2L_2(x)+\ln x\ln(1-x) \nn \\
\smvs &&\hspace{-0mm}
-\,\frac{(13-24x-27x^2+2x^3)}{2(1-x)^2}\ln(1-x) +
\frac{(12-22x-27x^2+2x^3)}{2(1-x)^2}\ln x \nn \\
\smvs &&\hspace{-0mm}
+\,3\frac{(4-12x+x^2+3x^3)}{2(1-x)^3}\ln\frac{\mu_m^2}{m_c^2} +
\frac{(6-64x+15x^2+11x^3)}{4(1-x)^3}\,\biggr\} \,.
\end{eqnarray}

The three-loop, order $\alpha_s^2$ correction $\rho^{(2)}(s)$ has
been calculated by Chetyrkin and
Steinhauser~\cite{Chetyrkin:2000mq,Chetyrkin:2001je} for the case of
one heavy and one massless quark. In the present analysis, we shall
make use of the program {\em Rvs.m}, which contains the required
expressions for
$\rho^{(2)}(s)$~\cite{Chetyrkin:2000mq,Chetyrkin:2001je}. However,
since the spectral function has been calculated only in the pole
mass scheme, following Jamin and Lange~\cite{Jamin:2001fw}, in the
$\MSb$ scheme we still have to add to $\rho^{(2)}(s)$ the
contributions resulting from rewriting the pole mass in terms of the
$\MSb$ mass. The two contributions $\Delta_1\rho^{(2)}$ and
$\Delta_2\rho^{(2)}$ which arise from the leading and first order
contributions, respectively, are given by
\begin{eqnarray}\label{del1rho2}
\Delta_1\rho^{(2)}(s) &=& \frac{N_c}{8\pi^2}\, (m_c+m_s)^2\,s\,
\Big[\,(3-20x+21x^2)\,r_m^{(1)^2} - 2(1-x)(1-3x)\,r_m^{(2)}\,\Big] \,, \\
\vbox{\vskip 9mm} \label{del2rho2} \Delta_2\rho^{(2)}(s) &\!=\!&
-\,\frac{N_c}{8\pi^2}\,C_F\, (m_c+m_s)^2\,s\,
r_m^{(1)}\biggl\{(1-x)(1-3x)\Big[\,4L_2(x)+2\ln x\ln(1-x)\,\Big] \nn \\
\vbox{\vskip 6mm}
&&\hspace{-0mm} -\,(1-x)(7-21x+8x^2)\ln(1-x) + (3-22x+29x^2-8x^3)\ln x \nn \\
\vbox{\vskip 9mm} &&\hspace{-0mm}
+\,\frac{1}{2}(1-x)(15-31x)\,\biggr\} \,,
\end{eqnarray}
where explicit expressions for the coefficients $r_m^{(1)}$ and
$r_m^{(2)}$ can be found in Appendix~B.

\subsection{The condensate contributions}

In the following, we summarize the contributions to the spectral
function $\rho_{\rm QCD}(s)$ coming from higher dimensional
operators, which arise in the framework of the OPE and parameterize
the appearance of non-perturbative physics. Since the spectral
functions corresponding to the condensates contain
$\delta$-distribution contributions, we shall present directly the
Borel transformed integrated quantity
$u\wh\Pi(u)=\int\limits_0^\infty e^{-s/u}\rho_{\rm QCD}(s)\,ds$
below, where $u=M_B^2$ with $M_B$ being the Borel mass.

The leading order expression for the dimension-three quark
condensate is well known with the explicit form given by
\begin{equation}\label{qq0}
u\wh\Pi^{(0)}_{\bar qq}(u) = -\,(m_c+m_s)^2 m_c\qq\,
e^{-m_c^2/u}\,\left[\, 1 - (1+\frac{m_c^2}{u})\frac{m_s}{2m_c} +
\frac{m_c^2 m_s^2}{2u^2} \,\right] \,,
\end{equation}
where the expansion up to order $m_s^2$ has been
included~\cite{Jamin:1992se}. The first order correction to the
quark condensate can be deduced based on the fact that the mass
logarithms must cancel once the quark condensate is expressed in
terms of the non-normal ordered
condensate~\cite{Jamin:1992se,Spiridonov:1988md,Chetyrkin:1994qu}
with
\begin{equation}\label{qq1}
u\wh\Pi^{(1)}_{\bar qq}(u) = \frac{3}{2}\,C_F\,a\,(m_c+m_s)^2
m_c\qq\, \biggl\{\Gamma(0,\frac{m_c^2}{u}) - \Big[\, 1 + (1 -
\frac{m_c^2}{u})\,(\ln\frac{\mu_m^2}{m_c^2}+\frac{4}{3})\,\Big]\,
e^{-m_c^2/u}\,\biggr\} \,,
\end{equation}
where $\Gamma(n,z)$ is the incomplete $\Gamma$-function.

The next contribution in the OPE is the dimension-four gluon
condensate with the corresponding expression given by
\begin{equation}\label{FF0}
u\wh\Pi^{(0)}_{FF}(u) \;=\;
\frac{1}{12}\,(m_c+m_s)^2\,\FF\,e^{-m_c^2/u} \,.
\end{equation}

The dimension-five mixed quark gluon condensate should also be
included, since it is enhanced by the heavy quark mass and hence
still has some influence on the sum rule. Again the result is well
known with
\begin{equation}\label{qFq0}
u\wh\Pi^{(0)}_{\bar qFq}(u) = -\,(m_c+m_s)^2\,\frac{m_c\langle
g_s\bar q\sigma Fq
\rangle}{2u}\,\Big(1-\frac{m_c^2}{2u}\Big)\,e^{-m_c^2/u} \,.
\end{equation}

The last condensate contribution considered in this paper is the
four-quark condensate
\begin{equation}\label{psibarpsi0}
u\wh\Pi^{(0)}_{(\bar s s)^2}(u) = -\sigma \frac{8\pi}{27}\,
\left(2-\frac{m_c^2}{2u}-\frac{m_c^4}{6u^2}\right)\, \alpha_s \la
\bar s s\ra^2,
\end{equation}
where $\sigma$ is the factor representing the deviation from vacuum
saturation. The contributions of all the other higher dimensional
operators are extremely small and thus have been neglected.

\section{Relationship between pole and running $\MSb$ quark mass}

The relationship between pole and running $\MSb$ quark mass is given
by~\cite{Jamin:2001fw}
\begin{equation}\label{eq:b.4}
m(\mu_m) \; = \; M_{{\rm pole}}\,\Big[\,1 +
a(\mu_a)\,r_m^{(1)}(\mu_m) + a(\mu_a)^2\,r_m^{(2)}(\mu_a,\mu_m) +
\ldots \,\Big] \,,
\end{equation}
where
\begin{eqnarray}
r_m^{(1)} & = & r_{m,0}^{(1)} - \gamma_1\ln\frac{\mu_m}{m(\mu_m)}
\,, \label{eq:b.5} \\ \smvs r_m^{(2)} & = & r_{m,0}^{(2)} -
\Big[\,\gamma_2+(\gamma_1-\beta_1)\,
r_{m,0}^{(1)}\,\Big]\ln\frac{\mu_m}{m(\mu_m)} +
\frac{\gamma_1}{2}\,(\gamma_1-
\beta_1)\ln^2\frac{\mu_m}{m(\mu_m)} \nn \\
\smvs & &
-\,\biggl[\,\gamma_1+\beta_1\ln\frac{\mu_m}{\mu_a}\,\biggr]
r_m^{(1)} \,. \label{eq:b.6}
\end{eqnarray}
The coefficients of the logarithms can be calculated from the
renormalisation group~\cite{Chetyrkin:1996cf}, and the constant
coefficients $r_{m,0}^{(1)}$ and $r_{m,0}^{(2)}$ are found to
be~\cite{Tarrach:1980up,Gray:1990yh}
\begin{eqnarray}
r_{m,0}^{(1)} & = & -\,C_F \,, \label{eq:b.7} \\
\smvs r_{m,0}^{(2)} & = &
C_F^2\biggl[\frac{7}{128}-\frac{15}{8}\zeta(2)-\frac{3}{4}
\zeta(3)+3\zeta(2)\ln
2\biggr]+C_FTn_f\biggl[\frac{71}{96}+\frac{1}{2}\zeta(2) \biggr] \nn
\\ \smvs & &
\hspace{-0mm}+\,C_AC_F\biggl[-\frac{1111}{384}+\frac{1}{2}\zeta(2)+
\frac{3}{8}\zeta(3)-\frac{3}{2}\zeta(2)\ln
2\biggr]+C_FT\biggl[\frac{3}{4}- \frac{3}{2}\zeta(2)\biggr] \,.
\label{eq:b.8}
\end{eqnarray}
with
\begin{equation}
\label{eq:b.2} \beta_1 \; = \;
\frac{1}{6}\,\Big[\,11C_A-4Tn_f\,\Big] \,, \qquad \beta_2 \; = \;
\frac{1}{12}\,\Big[\,17C_A^2-10C_ATn_f-6C_FTn_f\,\Big] \,,
\end{equation}
and
\begin{equation}
\label{eq:b.3} \gamma_1 \; = \; \frac{3}{2}\,C_F \,, \qquad \gamma_2
\; = \; \frac{C_F}{48}\,\Big[\,97C_A+9C_F-20Tn_f\,\Big] \,.
\end{equation}

\section{Relevant expressions in the $DK$ continuum contribution}

For convenience, in this appendix we collect some relevant
expressions used in Sec.~\ref{sec3} when discussing the $DK$
continuum contribution. Firstly, we define the following loop
integral functions $\Sigma_i(t)$~(with $t=q^2$)
\begin{eqnarray}\label{sigma0}
\Sigma_0(t) &=& 2i\int \frac{d^4k}{(2\pi)^4}\,
\frac{1}{(k^2-m_K^2)\, \big[(q-k)^2-m_D^2\big]} \nonumber \\
&=& -\frac{1}{8\pi^2}\,B_0(t,m_D^2,m_K^2)\,,
\end{eqnarray}
\begin{eqnarray}\label{sigma1}
\Sigma_1(t) &=& 2i\int \frac{d^4k}{(2\pi)^4}\,
\frac{k_0}{(k^2-m_K^2)\, \big[(q-k)^2-m_D^2\big]} \nonumber \\
&=&-\frac{1}{8\pi^2}\,\bigg\{\,\frac{t+m_K^2-m_D^2}{2\sqrt{t}}\,
B_0(t,m_D^2,m_K^2) +
\frac{1}{2\sqrt{t}}\,\Big[A_0(m_D^2)-A_0(m_K^2)\Big]\,\bigg\}\,,
\end{eqnarray}
\begin{eqnarray}\label{sigma2}
\Sigma_2(t) &=& 2i\int \frac{d^4k}{(2\pi)^4}\,
\frac{k_0^2}{(k^2-m_K^2)\, \big[(q-k)^2-m_D^2\big]} \nonumber \\
&=&-\frac{1}{8\pi^2}\,\bigg\{\,\frac{(t+m_K^2-m_D^2)^2}{4t}\,
B_0(t,m_D^2,m_K^2) +
\frac{t+m_K^2-m_D^2}{4t}\,\Big[A_0(m_D^2)-A_0(m_K^2)\Big] \nonumber \\
&& \hspace{5mm} +\frac{1}{2}\,A_0(m_D^2)\,\bigg\}\,,
\end{eqnarray}
\begin{eqnarray}\label{sigma3}
\Sigma_3(t) &=& 2i\int \frac{d^4k}{(2\pi)^4}\,
\frac{k^2}{(k^2-m_K^2)\, \Big[(q-k)^2-m_D^2\Big]} \nonumber \\
&=&-\frac{1}{8\pi^2}\,\Big[m_K^2\, B_0(t,m_D^2,m_K^2) +
A_0(m_D^2)\,\Big]\,,
\end{eqnarray}
\begin{eqnarray}\label{sigma4}
\Sigma_4(t) &=& 2i\int \frac{d^4k}{(2\pi)^4}\,
\frac{k^2 k_0}{(k^2-m_K^2)\, \big[(q-k)^2-m_D^2\big]} \nonumber \\
&=&-\frac{1}{8\pi^2}\,\bigg\{\,\frac{t+m_K^2-m_D^2}{2\sqrt{t}}\,
m_K^2\,B_0(t,m_D^2,m_K^2) + \frac{m_K^2}{2\sqrt{t}}\,
\Big[A_0(m_D^2)-A_0(m_K^2)\Big] \nonumber \\
&& \hspace{5mm} + \sqrt{t}\,A_0(m_D^2)\,\bigg\}\,,
\end{eqnarray}
\begin{eqnarray}\label{sigma5}
\Sigma_5(t) &=& 2i\int \frac{d^4k}{(2\pi)^4}\,
\frac{(k^2)^2}{(k^2-m_K^2)\, \Big[(q-k)^2-m_D^2\Big]} \nonumber \\
&=&-\frac{1}{8\pi^2}\,\Big[m_K^4\,B_0(t,m_D^2,m_K^2) +
(t+m_K^2+m_D^2)\,A_0(m_D^2)\,\Big]\,.
\end{eqnarray}
Here we have taken into account two intermediate states with
different charges in Eqs.~(\ref{sigma0})--(\ref{sigma5}). $A_0(m^2)$
and $B_0(t,m_1^2,m_2^2)$ is the usual one-loop scalar one- and
two-point function, respectively~\cite{Passarino:1978jh}
\begin{eqnarray}
A_0(m^2) &=& -i \int \frac{d^4k}{\pi^2}\, \frac{1}{(k^2-m^2)} \nonumber \\
&=&
m^2\Big[(\frac{2}{\epsilon}-\gamma_E+\ln{4\pi})+1-\ln{m^2\over\mu^2}\Big]\,,
\end{eqnarray}
\begin{eqnarray}
B_0(t,m_1^2,m_2^2) &=& -i \int \frac{d^4 k}{\pi^2}\,
\frac{1}{(k^2-m_1^2)\,\big[(q^2-k)^2-m_2^2\big]} \nonumber \\
&=& (\frac{2}{\epsilon}-\gamma_E+\ln{4\pi})+\ln{\mu^2} -
F_0(t,m_1^2,m_2^2)\,,
\end{eqnarray}
where $\epsilon=4-D$ in $D$-dimensional space time, $\mu$ is the
introduced renormalization scale in dimensional regularization, and
the explicit form of the function $F_0(t,m_1^2,m_2^2)$ could be
found in Ref.~\cite{Drees:1991rd}.

From the three linear equations for the three unknown functions
$f_i(t)$ given by Eqs.~(\ref{eq1})--(\ref{eq3}), we can deduce the
explicit expressions for the three unknown functions $f_i(t)$
\begin{eqnarray}\label{fit}
f_0(t) &=& \frac{1}{Y(t)}\,4 f_K^4 g_0 \left[\Sigma_0
f_K^2+\Sigma_1^2-\Sigma_0 \Sigma_2\right] (t-t_0)\,, \\
f_1(t) &=& \frac{1}{Y(t)}\, 2 f_K^4 g_0 \left[2 \Sigma_1
f_K^2-\Sigma_1 \Sigma_3+\Sigma_0 \Sigma_4+2 \left(\Sigma_1^2 -
\Sigma_0 \Sigma_2\right) \sqrt{t}\right] (t-t_0)\,, \\
f_2(t) &=& \frac{1}{Y(t)}\, 2 f_K^2 g_0 \left\{2 \Sigma_3
f_K^4+\big[2 \Sigma_1 \Sigma_4-\Sigma_3 (2
\Sigma_2+\Sigma_3)+\Sigma_0 \Sigma_5+2 (\Sigma_1 \Sigma_3-\Sigma_0
\Sigma_4) \sqrt{t}\,\big] f_K^2\, \right. \nonumber \\
&& \left. \hspace*{1cm} +\Sigma_0 \Sigma_4^2-2 \Sigma_1 \Sigma_3
\Sigma_4+\Sigma_1^2 \Sigma_5+\Sigma_2\left(\Sigma_3^2-\Sigma_0
\Sigma_5\right)\right\} (t-t_0)\,,
\end{eqnarray}
with
\begin{eqnarray}
Y(t) &=& \biggl\{4 \left(\Sigma_0 g_0^2+M_0^2-t\right) f_K^6+4
\left[\left(\Sigma_1^2-\Sigma_0 \Sigma_2\right)
g_0^2+\left(\Sigma_2+\Sigma_3-2\Sigma_1 \sqrt{t}\,\right)
\left(t-M_0^2\right)\right] f_K^4\, \nonumber \\
&& \hspace{-5mm} +\left(M_0^2-t\right) \left[4 t \Sigma_1^2-4
\Sigma_4 \Sigma_1+\Sigma_3^2-\Sigma_0 \Sigma_5+4 \Sigma_2
(\Sigma_3-\Sigma_0 t)-4 (\Sigma_1 \Sigma_3-\Sigma_0 \Sigma_4)
\sqrt{t}\right] f_K^2\, \nonumber \\
&& \hspace{-5mm} +\Big[2 \Sigma_3 \Sigma_4 \Sigma_1-\Sigma_5
\Sigma_1^2-\Sigma_0 \Sigma_4^2-\Sigma_2 \left(\Sigma_3^2-\Sigma_0
\Sigma_5\right)\Big] \left(M_0^2-t\right)\biggr\} (t-t_0)\, \nonumber \\
&& \hspace{-5mm} -4 c f_K^4 g_0^2 \left(\Sigma_0
f_K^2+\Sigma_1^2-\Sigma_0 \Sigma_2\right) \left(M_0^2-t\right)\,.
\end{eqnarray}

With the above results, the functions $\Delta(t)$ and $\Delta_1(t)$
can be, respectively, written as
\begin{eqnarray}\label{deltat}
\Delta(t) &=& {{(3t-m_D^2+m_K^2)(M_0^2-t)} \over {32 f_K^2 \pi^2
t}}\times A_0(m_D^2) + {{(3t-m_K^2+m_D^2)(M_0^2-t)} \over {32 f_K^2
\pi^2 t}}\times A_0(m_K^2) \nonumber \\
& & + {{c f_K^2 g_0^2 (t-M_0^2) + (f_K^2 g_0^2 -
(t-M_0^2)(t-m_D^2))(t-t_0)} \over {256 f_K^4 \pi^4 t (t-t_0)}}\times
A_0(m_D^2)^2 \nonumber \\
& & + {{c f_K^2 g_0^2 (t-M_0^2) + (f_K^2 g_0^2 -
(t-M_0^2)(t-m_K^2))(t-t_0)} \over {256 f_K^4 \pi^4 t (t-t_0)}}\times
A_0(m_K^2)^2 \nonumber \\
& & - {{2 c f_K^2 g_0^2 (t-M_0^2) + (2 f_K^2 g_0^2 +
(t-M_0^2)(t+m_D^2+m_K^2))(t-t_0)} \over {256 f_K^4 \pi^4 t
(t-t_0)}}\times A_0(m_D^2)A_0(m_K^2) \nonumber \\
&& - {{(t-m_D^2+m_K^2)A_0(m_D^2)+(t-m_K^2+m_D^2)A_0(m_K^2)} \over
{8192 f_K^6 \pi^6 t}}\times (t-M_0^2) A_0(m_D^2) A_0(m_K^2) \nonumber \\
&& - {{g_0^2} \over {8 \pi^2}} \times B_0(t,m_D^2,m_K^2) - {{c g_0^2
(t-M_0^2)} \over {8 \pi^2 (t-t_0)}} \times B_0(t,m_D^2,m_K^2) \nonumber \\
&& - {{(t-M_0^2)((m_D^2-m_K^2)^2 +2 (m_D^2+m_K^2)t
-3 t^2)} \over {32 f_K^2 \pi^2 t}} \times B_0(t,m_D^2,m_K^2) \nonumber \\
&& - {{(t+m_D^2-m_K^2)(c f_K^2 g_0^2 (t-M_0^2) + (f_K^2 g_0^2
-(t-M_0^2)(t-m_D^2))(t-t_0))} \over {256 f_K^4 \pi^4 t (t-t_0)}} \nonumber \\
&& \times A_0(m_D^2) B_0(t,m_D^2,m_K^2) \nonumber \\
&& - {{(t+m_K^2-m_D^2)(c f_K^2 g_0^2 (t-M_0^2) + (f_K^2 g_0^2
-(t-M_0^2)(t-m_K^2))(t-t_0))} \over {256 f_K^4 \pi^4 t (t-t_0)}} \nonumber \\
&& \times A_0(m_K^2) B_0(t,m_D^2,m_K^2) \nonumber \\
&& + {{(t^2-(m_D^2-m_K^2)^2)(t-M_0^2)} \over {8192 f_K^6 \pi^6 t}}
\times A_0(m_D^2) A_0(m_K^2) B_0(t,m_D^2,m_K^2)\,.
\end{eqnarray}
\begin{equation}\label{delta1t}
\Delta_1(t)=N(t)/D(t)\,,
\end{equation}
\begin{eqnarray}\label{nt}
N(t) &\; = \;& 32f_K^4\,g_0^2\,\pi^2\,(t-t_0)\,\left[A_0(m_K^2)^2 +
A_0(m_D^2)^2\right] - 64f_K^4\,g_0^2\,\pi^2\,(t-t_0)\,A_0(m_K^2)\,A_0(m_D^2) \nn \\
&& \hspace{-3mm}
-32f_K^4\,g_0^2\,\pi^2\,(t-t_0)\left[\,(m_D^2-m_K^2+t)\,A_0(m_D^2)
-(m_D^2-m_K^2-t)\,A_0(m_K^2)\,\right] \nn \\
&& \hspace{-3mm} \times \,B_0(t,m_D^2,m_K^2) -
1024f_K^6\,g_0^2\,\pi^4\,t\,(t-t_0)\,B_0(t,m_D^2,m_K^2) \,,
\end{eqnarray}
\begin{eqnarray}\label{dt}
D(t) &=& 256f_K^4\,\pi^4\,(t - t_0)\,\left[\,(3t -m_D^2 +
m_K^2)\, A_0(m_D^2) + (3t + m_D^2 - m_K^2)\, A_0(m_K^2)\, \right] \nn \\
&& \hspace{-3mm} + (t - t_0)\,\left[\,( t - m_D^2 + m_K^2)\,
A_0(m_D^2) + (t + m_D^2 - m_K^2)\, A_0(m_K^2)\, \right]\,
A_0(m_D^2)\,A_0(m_K^2)\, \nn \\
&& \hspace{-3mm} + 32f_K^2\,\pi^2\,\left[\,2c\,f_K^2\,g_0^2 + (m_D^2
+ m_K^2 + t)\, (t - t_0)\,\right] A_0(m_D^2)\, A_0(m_K^2) -
32f_K^2\,\pi^2\, \left\{\, \right. \nn \\
&& \hspace{-3mm} \left. \times \left[c\,f_K^2\,g_0^2 + (m_D^2 -
t)\,(t - t_0) \right]\, A_0(m_D^2)^2 + \left[c\,f_K^2\,g_0^2 +
(m_K^2 - t)\,(t - t_0) \right]\, A_0(m_K^2)^2\, \right\}\, \nn \\
&& \hspace{-3mm} + \left[\,(m_D^2 - m_K^2)^2 -
t^2\,\right]\,(t - t_0)\,
A_0(m_D^2)\, A_0(m_K^2)\, B_0(t,m_D^2,m_K^2) \nn \\
&& \hspace{-3mm} + 32f_K^2\,\pi^2\,\left\{(m_D^2 - m_K^2 +
t)\,\left[c\,f_K^2\,g_0^2 + (m_D^2 - t)\,(t - t_0)\,\right]\,
A_0(m_D^2) \right. \nn \\
&& \hspace{-3mm} \left. - (m_D^2 - m_K^2 - t)\,\left[c\,f_K^2\,g_0^2
+ (m_K^2 - t)\,(t - t_0)\,\right]\, A_0(m_K^2) \right\}
\,B_0(t,m_D^2,m_K^2)\, \nn \\
&& \hspace{-3mm} + 256f_K^4\,\pi^4\,
\left\{\,4c\,f_K^2\,g_0^2\,t + \left[(m_D^2 - m_K^2)^2 + 2( m_D^2 +
m_K^2)\,t - 3t^2 \right]\,(t - t_0) \right\}\, \nn \\
&& \hspace{-3mm} \times B_0(t,m_D^2,m_K^2) + 8192f_K^6\,\pi^6\, t\,
(t-t_0)\,.
\end{eqnarray}

\end{document}